\documentclass[%
 reprint,
 floatfix,
 twocolumn,
 amsmath,amssymb,
 aps,
]{revtex4-1}

\usepackage{amsmath}
\usepackage{graphicx}% Include figure files
\usepackage{dcolumn}% Align table columns on decimal point
\usepackage{bm}% bold math
\usepackage{braket}
\usepackage{xcolor}
\usepackage{ulem}

\begin{document}

\preprint{APS/123-QED}

\title{Orbital Hall physics in two-dimensional Dirac materials}% Force line breaks with \\

\author{Armando Pezo}
\email{armando-arquimedes.pezo-lopez@univ-amu.fr}
 \affiliation{Aix-Marseille Universit\'e, CNRS, CINaM, Marseille, France.}
\author{Diego Garc\'ia Ovalle}
%\email{email: ovalle@cinam.univ-mrs.fr}
 \affiliation{Aix-Marseille Universit\'e, CNRS, CINaM, Marseille, France.}
\author{Aur\'elien Manchon}%
\email{aurelien.manchon@univ-amu.fr}
\affiliation{Aix-Marseille Universit\'e, CNRS, CINaM, Marseille, France.}

%\email{email: email: %ovalle@cinam.univ-mrs.fr.cl}

%\affiliation{Aix-Marseille Univiversité, CNRS, CINaM, Marseille, France.} 
\date{\today}

\begin{abstract}
Orbitronics has recently emerged as a very active research topic after several proposals aiming to exploit the orbital degree of freedom for charge-free electronics. In this communication, we investigate orbital transport in selected two-dimensional systems to better understand which parameters govern the intra-atomic and inter-atomic contributions to the orbital Hall effect. We study the impact of the gap, the role of the materials' topology and the influence of the disorder on spin and orbital Hall transport. Starting from the Kane-Mele model, we describe how the orbital moment behaves depending on the material's topology and clarify the influence of the gap on the orbital Hall conductivity. We then extend the study to realistic topologically trivial and non-trivial materials, and find that the topology has little qualitative influence on the orbital Hall conductivity. In contrast, we observe that the energy dispersion has a more dramatic impact, especially in the presence of disorder. Remarkably, our results suggest that the intra-atomic orbital Hall current is more robust against scattering than the inter-atomic one, without further impact of the topological properties of the system under consideration.
\end{abstract}

                              
\maketitle

%\tableofcontents
\noindent
\section{Introduction}
Recent theoretical and experimental efforts suggest that the orbital angular momentum of electrons can be used as an alternative degree of freedom to the spin angular momentum \cite{Bernevig2005b,Kontani2009,Go2018,Hayashi2012,Salemi2022,Pezo2022}. In contrast with the generation of spin currents that necessitates either a ferromagnet or a heavy metal, orbital currents can be induced electrically using light metals, thereby presenting a potential technical advantage in terms of materials scarcity \cite{Bernevig2005b,Go2018}. Current research is being developed along two directions. A first direction takes advantage of the vast experience acquired on spin transport in transition metal heterostructures \cite{Vedmedenko2020,Chumak2015}. Tight-binding and first principles calculations have suggested that certain light metals such as V, Cr or Cu can host large orbital Hall effect \cite{Jo2018,Salemi2022}, resulting in the experimental demonstration of orbital torque and magnetoresistance \cite{go2020,Ding2020,Lee2021b,Ding2022}. In the last years new experimental developments have unlocked the synthesis of two-dimensional materials opening new possibilities for spintronics \cite{Roche2015}. To date, most of the attention has been focused on graphene and transition metal dichalcogenides, where valley Hall effect and orbital Hall effect coexist \cite{Xiao2012b,Bhowal2021,Cysne2021,Cysne2022}.

In most early theoretical studies on the orbital Hall effect, the orbital moment was assumed to be mostly of intra-atomic origin, adopting the so-called atom-centered approximation (ACA) \cite{Bernevig2005b,Kontani2009,Jo2018,Canonico2020,Cysne2021,Salemi2022}. However, important developments in the theory of orbital magnetism have demonstrated that ACA is not sufficient to properly describe the orbital motion of quasiparticles in solids and that the inter-atomic contribution cannot be neglected \cite{Thonhauser2005,Ceresoli2006,Shi2007,Hanke2016}. Whereas the inter-atomic orbital moment is small in the case of bulk transition metals \cite{Hanke2016}, it is particularly significant in materials like graphene where the intra-atomic orbital character of the conduction electrons vanishes \cite{Liu2021_orbital_mag_graphene}. Extending the "modern theory" of orbital magnetization to the orbital Hall effect, it recently appeared that the inter-atomic contribution cannot be neglected in general \cite{Bhowal2021,Cysne2022,Pezo2022}. 

Building on our previous work on the modern theory of orbital Hall effect in realistic materials \cite{Pezo2022}, we investigate orbital transport in selected two-dimensional systems to better understand which parameters govern the intra-atomic and inter-atomic contributions to the orbital Hall effect in these systems. In particular, we investigate the impact of the gap, the role of the materials topology and the influence of the disorder on spin and orbital Hall transport. We find that whereas the topology has little influence on the orbital Hall effect itself, the orbital transport exhibits markedly distinct behavior depending on the nature of the gap (spin-orbit coupling, staggered potential) and systematically increases upon reducing the gap size. Our results also show that the intra-atomic and inter-atomic orbital Hall conductivities experience different robustness against disorder. Whereas the inter-atomic orbital Hall contribution systematically decreases upon increasing impurity scattering, the intra-atomic contribution remains mostly unaffected in the gap region and decreases in the metallic regime, suggesting that overall the intra-atomic contribution is more robust against disorder than the inter-atomic one.

This work is organized as follows. In Section \ref{s:1}, we briefly remind the concepts of the modern theory of the orbital Hall effect and use the two-dimensional Kane-Mele model to determine the influence of the lattice topology on the spin and orbital Hall transport. In Section \ref{s:2}, we investigate the spin and orbital Hall effects in selected two-dimensional lattices computed using density functional theory (DFT) and show that orbital Hall transport is substantially influenced by the proximity to the gap and by the nature of the energy dispersion. Then, in Section \ref{s:3}, we investigate the impact of the Anderson-type disorder on the orbital Hall effect and show that intra-atomic and inter-atomic contributions behave differently. Conclusions and perspectives are given in Section \ref{s:4}.

\section{Theory and concepts\label{s:1}}
\subsection{Modern theory of the orbital Hall effect}
Let us remind the concepts related to the modern theory of orbital magnetization and its extension to the orbital Hall effect. In crystals, the orbital motion of an electron arises from the self-rotation of the wavepacket in the unit cell which includes both the intra-atomic and inter-atomic contributions as mentioned above. In equilibrium, this self-rotation gives rise to the orbital magnetization as long as the time-reversal symmetry is broken \cite{Thonhauser2005,Shi2007}. In contrast, for materials with time-reversal symmetry, an orbital magnetization can be generated out of equilibrium as long as inversion symmetry is broken \cite{Yoda2018}. When both time-reversal symmetry and inversion symmetry are preserved though, no orbital magnetization can be induced, and only the orbital Hall effect survives.%Go2017

Let us start with the real space definition of the orbital moment, $\hat{\bf L}=(\mathbf{\hat{r}}\times \mathbf{\hat{p}}-\mathbf{\hat{p}}\times \mathbf{\hat{r}})/4$. Under the parallel transport gauge condition applied on non-degenerate bands (i.e., $\langle n|\dot{n}\rangle=0$), this expression can be projected on the Bloch states $|u^n_{\bf k}\rangle$ and recasted in the form (see \cite{Bhowal2021,Blount1962})

\begin{eqnarray}
\langle u^n_{\bf k}|\hat{\bf L}|u^p_{\bf k}\rangle=&\frac{e}{2g_L\mu_B}{\rm Im}\langle \partial_{\bf k}u^n_{\bf k}|\times\mathcal{H}_{\bf k}|\partial_{\bf k}u^p_{\bf k}\rangle\nonumber\\
&-\frac{e}{4g_L\mu_B}(\varepsilon^n_{\bf k}+\varepsilon^p_{\bf k}){\rm Im}\langle \partial_{\bf k}u^n_{\bf k}|\times|\partial_{\bf k}u^p_{\bf k}\rangle.
\end{eqnarray}

Here, $|u^n_{\bf k}\rangle$ is the periodic part of the Bloch state associated with the energy $\varepsilon^n_{\bf k}$, $\mathbf{\hat{v}}=\hbar^{-1}\partial_{\mathbf{k}} \mathcal{H}_{\mathbf{k}}$ is the velocity operator, $\mathcal{H}_{\mathbf{k}}$ being the Hamiltonian in momentum space, $\mu_B=e\hbar/2m_e$ is Bohr's magneton and $g_L=1$ is Land\'e's g-factor. This expression can be further formulated in a more tractable identity given by

\begin{align}
     \langle u^n_{\bf k}|\hat{\bf L}|u^p_{\bf k}\rangle&=\frac{e\hbar^2}{4\mu_B}\operatorname{Im}\sum_{q\not=n,p}\left(\frac{1}{\varepsilon^q_{\bf k}-\varepsilon^n_{\bf k}}+\frac{1}{\varepsilon^q_{\bf k}-\varepsilon^p_{\bf k}}\right)\nonumber\\
     &\braket{u^n_{\bf k}|\mathbf{\hat{v}}|u^q_{\bf k}}\times\braket{u^q_{\bf k}|\mathbf{\hat{v}}|u^p_{\bf k}}.\label{i3}
\end{align}

In the Bloch state basis, the matrix element of the orbital current operator, defined as $\mathcal{J}_i^\gamma=1/2\{L_\gamma,v_i\}$, reads

\begin{eqnarray}
\braket{u^n_{\bf k}|\mathcal{J}_i^\gamma|u^m_{\bf k}}=\frac{1}{2}\sum_p&\left(\langle u^n_{\bf k}|\hat{v}_i|u^p_{\bf k}\rangle\langle u^p_{\bf k}|L_\gamma|u^m_{\bf k}\rangle\right.\nonumber\\
&\left.+\langle u^n_{\bf k}|L_\gamma|u^p_{\bf k}\rangle\langle u^p_{\bf k}|\hat{v}_i|u^m_{\bf k}\rangle\right),\label{eq3}
\end{eqnarray}

where the orbital moment $L_{\gamma}$ is either the atomic orbital moment operator (intra-atomic orbital current) or Eq. \eqref{i3} (total orbital current). In the following, we will take $L_\gamma=L_z$ since we focus on two-dimensional lattices. In the linear response theory, spin and orbital currents are time-reversal symmetric which allows one to consider the intrinsic Fermi sea contribution of the Kubo formula \cite{Bonbien2020}. The orbital conductivity reads

\begin{equation*}
        \sigma_{ij}^{z}=-2\hbar e\int_{BZ}\frac{d^3{\bf k}}{(2\pi)^3}\sum_{n}f_n(\mathbf{k}) \times
\end{equation*}
\begin{equation}
        \operatorname{Im}\sum_{m\not=n} \frac{\braket{u^n_{\bf k}|\mathcal{J}_{i}^z|u^m_{\bf k}}\braket{u^m_{\bf k}|\hat{v}_j|u^n_{\bf k}}}{(\varepsilon^n_{\bf k}-\varepsilon^m_{\bf k})^2},\label{i1}
\end{equation}
where $f_n(\mathbf{k})$ is the equilibrium Fermi distribution function. The quantity that multiplies the Fermi function is usually referred to as the orbital Berry curvature, in analogy with the conventional Berry curvature where the orbital current operator is replaced by the velocity operator \cite{Cysne2021,Bhowal2020,Go2018}.

\subsection{Orbital Hall effect in Kane-Mele model}
\begin{figure}[ht!]
\includegraphics[width=\linewidth]{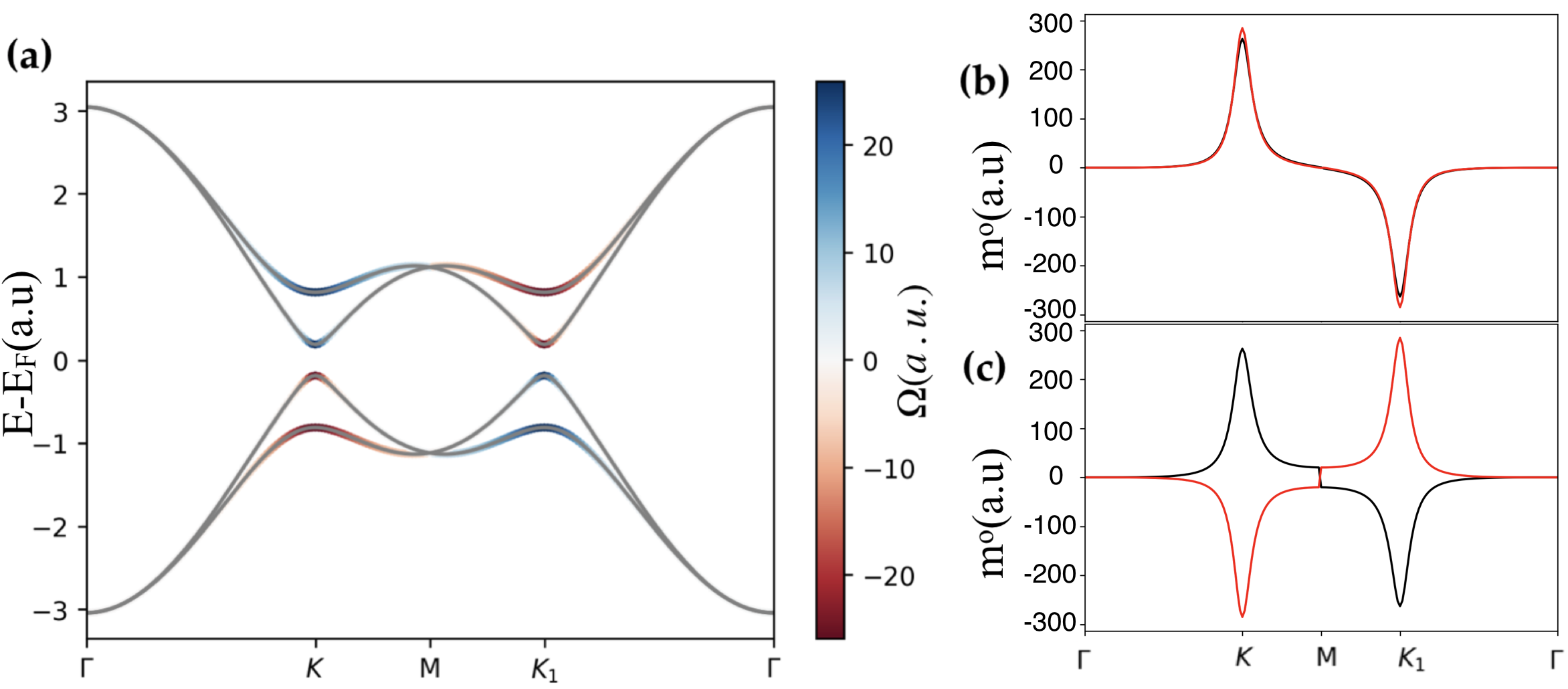}% %Here is how to import EPS art
\caption{Kane-Mele model band structure for the trivial phase with the Berry curvature displayed in the color bar (a). The orbital moment plotted for trivial ($\Delta=0.25t,\;\lambda_{\rm SOC}=0$) (b) and topological ($\Delta=0,\;\lambda_{\rm SOC}=0.25t$) (c) phases for the two valence energy bands. The black (red) curve refers to the lowest (highest) valence band.}
\label{fig:kane_mele}
\end{figure}

\begin{figure*}[ht]
\includegraphics[width=0.9\linewidth]{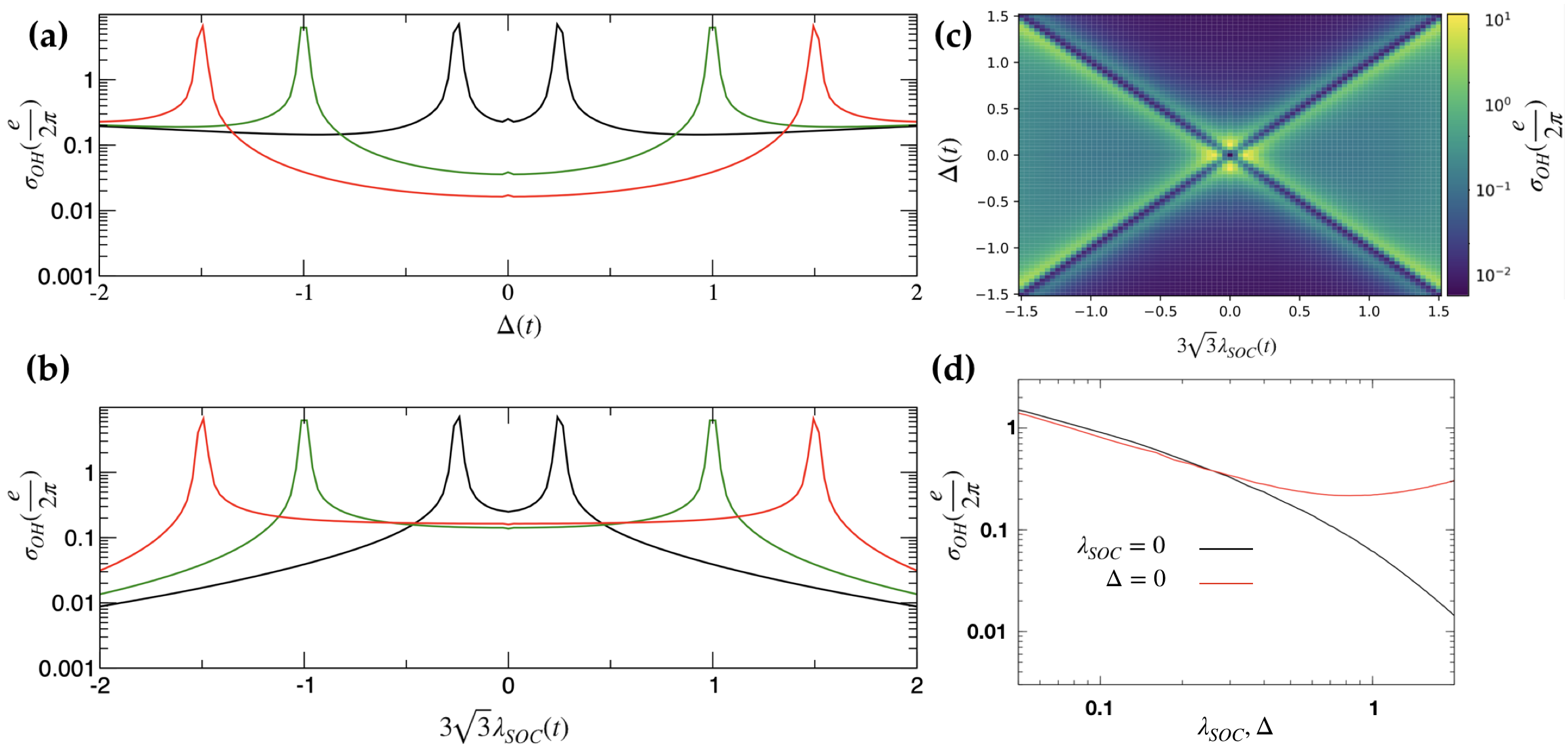}
\caption{(Color online) (a) Orbital Hall conductivity of the Kane-Mele model as function of $\Delta$ for fixed values of the $\lambda_{\rm SOC}$ [0.25$t$ (black), 1.0$t$ (red) and 1.5$t$ (green)]. (b) Orbital Hall conductivity as function of $\lambda_{\rm SOC}$ with fixed values of $\Delta$ [0.25$t$ (black), 1.0$t$ (red) and 1.5$t$ (green)]. (c) Phase diagram for the orbital Hall conductivity as function of both $\Delta$ and $\lambda_{\rm SOC}$. Whenever $|3\sqrt{3}\lambda_{\rm SOC}|>|\Delta|$, the orbital Hall effect is larger than for the opposite situation. (d) Orbital Hall conductivity in the center of the band gap as a function of $\Delta$ (black) and $\lambda_{\rm SOC}$ (red).}
\label{fig:ohc_phase}
\end{figure*}

Due to the ubiquitous presence of spin-orbit coupling in real materials, orbital currents usually coexist with spin currents and probing orbital-only responses is experimentally challenging. To overcome this difficulty, one usually relies on materials with vanishing or low spin-orbit coupling so that spin currents can be neglected \cite{Ding2020,Ding2021a}. In topological materials though, the topological transition is mostly driven by the strong spin-orbit coupling of the heavy elements, so that spin and orbital currents are entangled. A pedagogical tool to evaluate how these two effects behave in two-dimensional topological materials is the Kane-Mele Hamiltonian \cite{Kane2005}. This model features a two-dimensional honeycomb lattice with spin-orbit coupling and its Hamiltonian reads
\begin{equation}\label{km}
\mathcal{H}=t\sum_{\braket{ij}}c^{\dagger}_ic_j+i\lambda_{\rm SOC}\sum_{\braket{\braket{ij}}}\nu_{ij}c^{\dagger}_is_zc_j+\Delta \sum_i\epsilon_ic^{\dagger}_ic_i,
\end{equation}

where the creation and annihilation operators $c_i,\;c_i^{\dagger}$ are spinors representing the spin degree of freedom. The first term is the usual nearest neighbour hopping ($t$), the second term is the spin-orbit coupling term ($\lambda_{\rm SOC}$) acting on the next-nearest neighbours and the last term is the on-site staggered potential term ($\epsilon_i\Delta,\;\epsilon_i=\pm1$ on the different sublattices), also called Semenoff mass gap \cite{semenoff_1984}. This last term leads to the appearance of an orbital magnetic moment as pointed out in Refs. \cite{Bhowal2021,Bhowal2020}. To illustrate the connection between the Berry curvature, the orbital magnetic moment and the orbital Hall effect, let us consider the effective Hamiltonian of the Kane-Mele model, valid close to the neutrality point at K and K' points in the Brillouin zone. In the absence of Rashba interaction, the Kane-Mele model can be thought of as a double copy of the Haldane model \cite{Haldane1988}, each copy corresponding to a spin sector described by the two-band Hamiltonian
\begin{equation}
\mathcal{H}^{\eta s} =v_F(\eta k_x\sigma_x+k_y\sigma_y)+(\Delta+\eta s\tilde{\lambda}_{\rm SOC})\sigma_z,
\label{eq:trivial_ham}
\end{equation}
where $\tilde{\lambda}_{\rm SOC}=3\sqrt{3} \lambda_{\rm SOC}$, ${\bm \sigma}$ is the pseudospin in the sublattice space, $s=\pm1$ refers to the spin projection and $\eta=\pm1$ to the K and K' valleys, respectively. The gap is given by $\Delta_{\eta s}=\Delta+\eta s\tilde{\lambda}_{\rm SOC}$ so that the non-trivial regime is reached whenever $|\tilde{\lambda}_{\rm SOC}|>|\Delta|$. This Hamiltonian can be written \cite{Yoda2018,Xiao_liang_topo_2006}
\begin{equation}
    \mathcal{H}^{\eta s}=\hat{I}\epsilon_k^{\eta s}+\hat{\bm\sigma} \cdot \mathbf{d_k^{\eta s}},
\end{equation}
where $\epsilon_k^{\eta s}$ is the energy dispersion of the individual bands and $\mathbf{d_k^{\eta s}}$ describes the hybridization between bands. Therefore the Berry curvature and orbital magnetic moment read \cite{Yoda2018}
\begin{eqnarray}
    \Omega_{k,i}^{\eta s}&=&\pm\frac{\varepsilon_{ijk}}{2(d_k^{\eta s})^3}\mathbf{d_k^{\eta s}}\cdot\left(\frac{\partial \mathbf{d_k^{\eta s}}}{\partial k_j}\times\frac{\partial \mathbf{d_k^{\eta s}}}{\partial k_k}\right),
    \label{bc}\\
    m_{k,i}^{o,\eta s}&=&-\frac{e}{\hbar}\frac{\varepsilon_{ijk}}{2(d_k^{\eta s})^2}\mathbf{d_k^{\eta s}}\cdot\left(\frac{\partial \mathbf{d_k^{\eta s}}}{\partial k_j}\times\frac{\partial \mathbf{d_k^{\eta s}}}{\partial k_k}\right),
    \label{om}
\end{eqnarray}
with $\pm$ referring to the conduction and valence band, respectively and $d_k^{\eta s}=|\mathbf{d_k^{\eta s}}|$. Equations \eqref{bc} and \eqref{om} show that the Berry curvature and the orbital magnetic moment possess a very similar structure in momentum space \cite{Bhowal2021,Thonhauser2005}, revealing that the orbital motion finds it origin in the Berry curvature of the Bloch states. Since the Kane-Mele model, Eq. \eqref{km}, describes an effective four-band model taking into account the spin degree of freedom, the Bloch state does not carry any atomic orbital momentum (in other words, in this model the Bloch states are typically $s$ or $p_z$). For this reason, the orbital moment comes entirely from the details of the band structure intimately correlated to the Berry curvature, as displayed as a color gradient in Fig. \ref{fig:kane_mele}(a). Using the effective two-band Hamiltonian, one obtains
\begin{eqnarray}
\Omega_{k,z}^{\eta s}=\pm\frac{\eta v_F^2\Delta_{\eta s}}{2(\Delta^2_{
\eta s}+v_F^2k^2)^{3/2}},\\
m_{k,z}^{o,\eta s}=-\frac{e}{2\hbar}\frac{\eta v_F^2\Delta_{\eta s}}{\Delta^2_{
\eta s}+v_F^2k^2}.
\end{eqnarray}
One sees that both the Berry curvature and the orbital magnetic moment are inversely proportional to the band gap \cite{Bhowal2021}. The Berry curvature hot-spots appearing at different valleys characterize the topological transition: in the non-trivial phase both spin partners at the same inequivalent point (K or K') display opposite values of the orbital moment, while they have the same sign in the trivial regime, as seen in Fig. \ref{fig:kane_mele}(b). 
%Using this argument and Fig. \ref{fig:kane_mele}(b) we proceed to make a concrete distinction between trivial and non-trivial phases. In gapped graphene, it is known that the orbital moment is proportionally inverse to the band gap \cite{Bhowal2021}, that is, in the model system considered in our study it is mostly determined by $\Delta$ and $\lambda_{\rm SOC}$. 

%These same parameters tune the phases of the Kane-Mele model for which the orbital moment is depicted in Fig. \ref{fig:kane_mele}(b) where we observe a qualitative difference; 

Following the procedure outlined by \citet{Bhowal2021}, the orbital Hall conductivity for spin $s$ and valley $\eta$ reads
\begin{equation}\label{eq:oh1}
\sigma_{OH}^{\eta s} = \left(\frac{e}{2\pi}\right)\left(\frac{m_ev_F^2}{6\hbar^2g_L}\right)\frac{\Delta^2_{\eta s}}{(\Delta^2_{
\eta s}+v_F^2k^2)^{3/2}},
\end{equation}
Therefore, the total orbital Hall conductivity is the sum of individuals contributions, $\sigma_{OH}=\sum_{\eta}\sigma_{OH}^{\eta}$. In the middle of the gap ($k\rightarrow0$), one retrieves the constant value pointed out by \citet{Bhowal2021},  
\begin{equation}\label{eq:oh2}
\sigma_{OH}= \left(\frac{e}{2\pi}\right)\left(\frac{m_ev_F^2}{3\hbar^2g_L}\right)\left(\frac{1}{|\Delta+\tilde{\lambda}_{\rm SOC}|}+\frac{1}{|\Delta-\tilde{\lambda}_{\rm SOC}|}\right).
\end{equation}
This expression indicates that the orbital Hall conductivity decreases when increasing the gap, independently of the topological nature of the gap. To assess the interplay between the spin-orbit coupling strength $\lambda_{\rm SOC}$ and the staggered potential $\Delta$, Fig. \ref{fig:ohc_phase}(a) [Fig. \ref{fig:ohc_phase}(b)] shows the orbital Hall conductivity as a function of $\Delta$ ($\lambda_{\rm SOC}$) for different values of $\lambda_{\rm SOC}$ ($\Delta$). We find that the orbital Hall conductivity reaches a maximum whenever $\Delta\approx\tilde{\lambda}_{\rm SOC}$, i.e., at the topological phase transition. We note that the orbital Hall conductivity tends to be larger in the trivial phase than in the topological phase, as confirmed by the phase diagram presented in Fig. \ref{fig:ohc_phase}(c). In this panel, brighter regions correspond to larger values of the orbital Hall conductivity, located in the topologically-trivial regions, $|\Delta|>|\tilde{\lambda}_{\rm SOC}|$. 

\begin{figure}[ht!]
\includegraphics[width=0.9\linewidth]{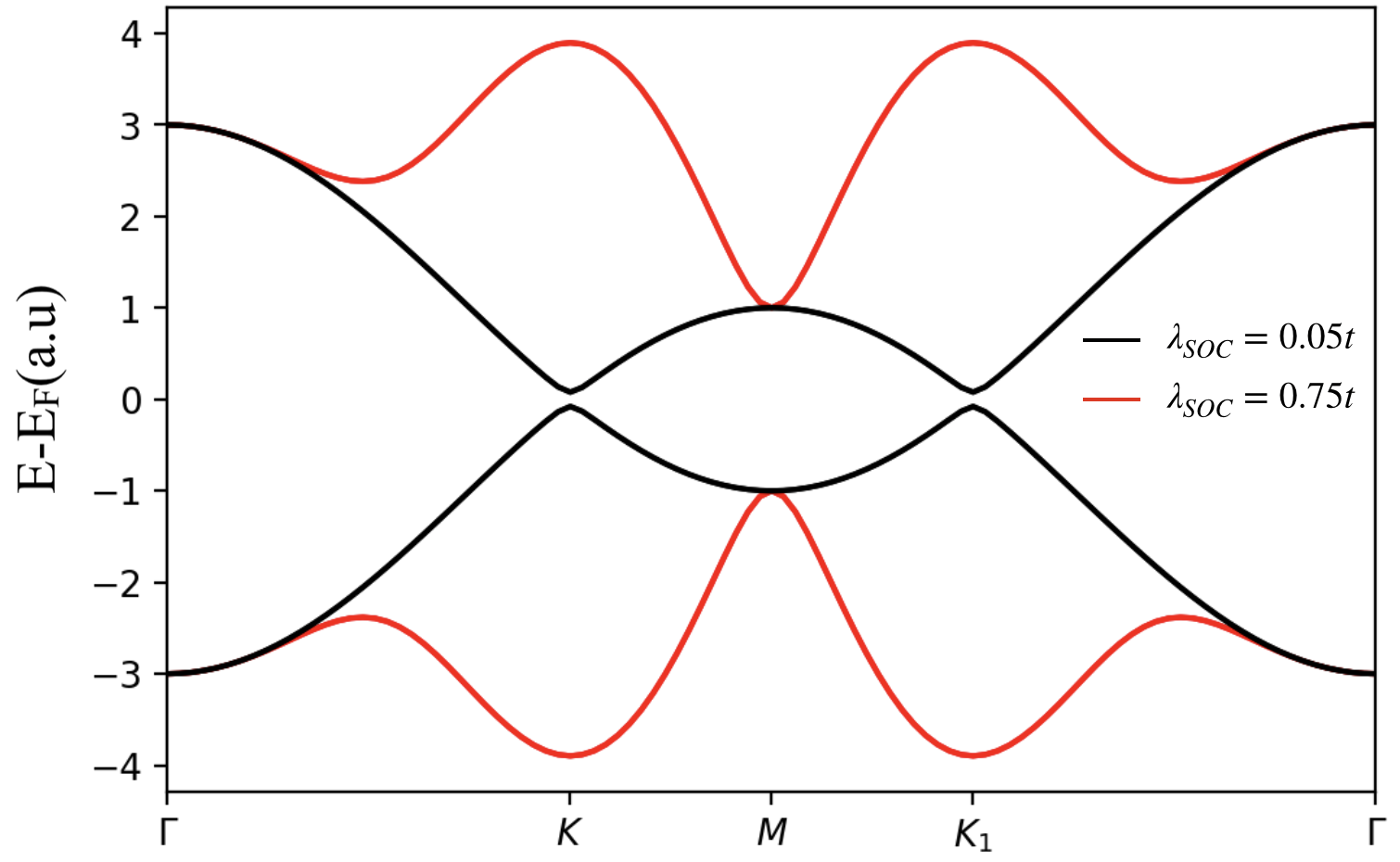}% %Here is how to import EPS art
\caption{Band structure for the Kane-Mele model with different values of $\lambda_{\rm SOC}$. In the non-trivial phase, the increasing of spin orbit coupling not necessarily leads to the increase of the gap, this is in connection to what is depicted in Fig. \ref{fig:ohc_phase}(d).}
\label{fig:kane_mele_soc}
\end{figure}

Finally, in order to establish a connection between the Kane-Mele model and the realistic two-dimensional materials discussed in the next section, we report the orbital Hall conductivity as a function of the gap size in Fig. \ref{fig:ohc_phase}(d). Our calculations confirm that the orbital Hall conductivity systematically decreases with the size for the gap, be it driven by $\Delta$ or by $\tilde{\lambda}_{\rm SOC}$ consistently with Eq. \eqref{eq:oh2}. When the staggered potential is turned off ($\Delta=0$, red curve), at small $\tilde{\lambda}_{\rm SOC}$, the gap is located close to the K and K' points, as shown in Fig. \ref{fig:kane_mele_soc}, and Eq. \eqref{eq:oh2} applies. However, in the large spin-orbit coupling limit, i.e., $\tilde{\lambda}_{\rm SOC}\approx t$, the gap has moved to the M point (Fig. \ref{fig:kane_mele_soc}). In this case, increasing the spin-orbit coupling strength maintains a gapped spectrum while incrementing the Fermi velocity, which leads to an enhancement of the orbital Hall effect, as depicted in Fig. \ref{fig:ohc_phase}(d). The two situations reported in Fig. \ref{fig:kane_mele_soc} are representative of the case of germanene (small $\tilde{\lambda}_{\rm SOC}$, Dirac cones at K and K' points) and bismuthene (large $\tilde{\lambda}_{\rm SOC}$, Dirac cone at $\Gamma$ point) discussed below.

\section{Orbital transport in realistic 2D materials\label{s:2}}

\begin{figure*}
\includegraphics[width=0.9\linewidth]{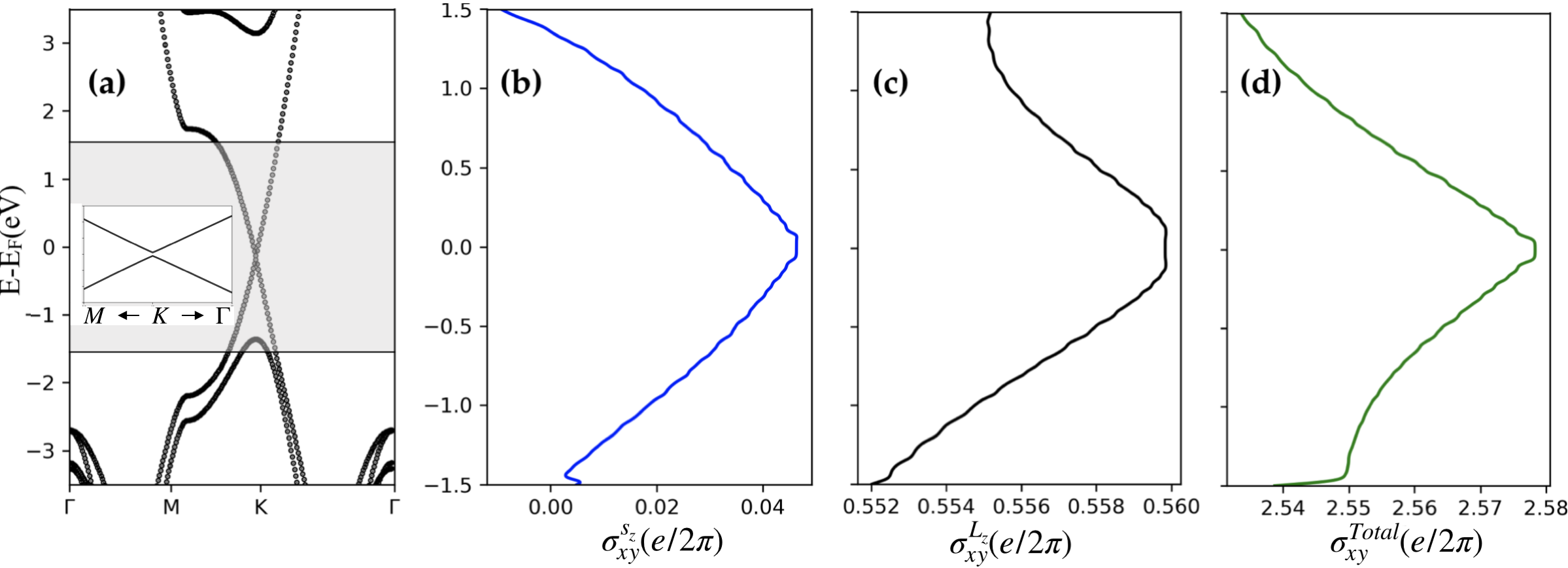}% %Here is how to import EPS art
\caption{(a) h-BN/graphene heterostructure electronic band structure showing the band gap opening at the Dirac point the inset, (b) spin Hall, (c) intra-atomic orbital Hall and (d) total orbital Hall conductivities. The grey shaded region in (a) corresponds to the energy window where the Hall conductivities were calculated.}
\label{fig:hbn_graphene_full}
\end{figure*}

We now turn to the simulations performed in realistic two-dimensional materials presenting different topological characters. For the trivial systems, we consider two cases: h-BN/graphene bilayer where the proximity effects lead to the breaking of inversion symmetry \cite{gra_hbn_C7RA00260B,graphene_hbn_soc_PhysRevB.103.075129}, as well as hydrogene-decorated graphene in which a colossal enhancement of the spin-orbit coupling has been predicted \cite{graphene_h_PhysRevLett.110.246602}. For the non-trivial systems, we have selected bismuthene, proven to be a topological insulator with a sizeable gap in its buckled hexagonal structure \cite{Guo2017b}, and germanene, characterized by a buckled structure and a large enough spin-orbit coupling capable to open a topological gap \cite{Acun2015,Ezawa2015}. For the DFT \cite{dft1964,dft1965} simulations, we used the Perdew-Burke-Ernzerhof \cite{gga,pbe} exchange-correlation functional. We performed the relaxation with the plane-wave basis as implemented in the Vienna $\textit{Ab-initio}$ Simulation Package (VASP) \cite{vasp1,vasp2}, and employ a plane-wave expansion cutoff  of 400 eV along with a force criterion of 0.2$\times10^{-2}$ eV/\AA\;with a $(15 \times 15 \times 1)$  $\mathbf{k}$-points sampling of the Brillouin zone. The ionic potentials were described using the projector augmented-wave (PAW) method \cite{paw}. Finally, the Hamiltonian matrix was obtained through the Wannier90 package.  The tight-binding representation was obtained by using a set of localized Wannier functions via the Wannier90/VASP interface \cite{Pizzi2020}. To do so, we first provide a trial set of functions which represent the actual atomic orbitals in the system under study leading to a Hamiltonian written in terms of orthogonal Wannier functions. The evaluation of any physical observable is performed from the obtention of the eigenenergies and eigenvectors of this Hamiltonian \cite{Pizzi2020}.

\subsubsection{h-BN/graphene and graphene+H}

The recent proposal suggesting gapped graphene as an orbital Hall insulator has shed light on the nature of the valley Hall effect \cite{Bhowal2021}. Motivated by this realization, we present the results on h-BN/graphene heterostructures. This system has been extensively studied in the last years in various contexts \cite{gra_hbn_C7RA00260B,graphene_hbN_correlated_Sun2021,hbn_graphene_non_local}. While free-standing graphene is a topological semimetal that possesses a robust band structure protected by inversion symmetry, it loses this symmetry by proximity with a material like h-BN. The whole heterostructure resembles a Kane-Mele model in the trivial phase with a gap opening whose size is given by the interaction with h-BN. Theoretically it has been shown that spin manipulation would be possible in this scenario \cite{graphene_hbn_soc_PhysRevB.103.075129}. Most importantly for our purpose, graphene acquires a $p_x$-$p_y$ orbital hybridization when interfaced with h-BN, which promotes the onset of intra-atomic orbital Hall effect. The Hall conductivities are shown in Fig. \ref{fig:hbn_graphene_full} where a small spin Hall effect is observed [Fig. \ref{fig:hbn_graphene_full}(b)], whereas the orbital Hall response is one to two orders of magnitude larger. In particular, the intra-atomic orbital Hall effect [Fig. \ref{fig:hbn_graphene_full}(c)] displays a moderate value within the energy window around the charge neutrality point while the total orbital Hall effect, which contains both intra- and inter-atomic contributions, [Fig. \ref{fig:hbn_graphene_full}(d)] attains the largest value of the three Hall responses. Notice that the energy profile of the Hall responses are similar, as both spin and orbital Hall effects are driven by proximity with h-BN. We point out that $d$ orbitals were previously introduced in order to increase the accuracy of the band structure compared with that obtained from GW+DFT simulations \cite{10.1063/1.3582136}. By enlarging the orbital basis to account for such $d$ orbitals, we found that their contribution to the orbital Hall conductivity is close to $\sim5$\%. This is rather negligible, especially in the present case where the largest contribution comes from the total orbital moment and is not related to the atomic orbital character of the Bloch state.

\begin{figure}[ht!]
\includegraphics[width=0.9\linewidth]{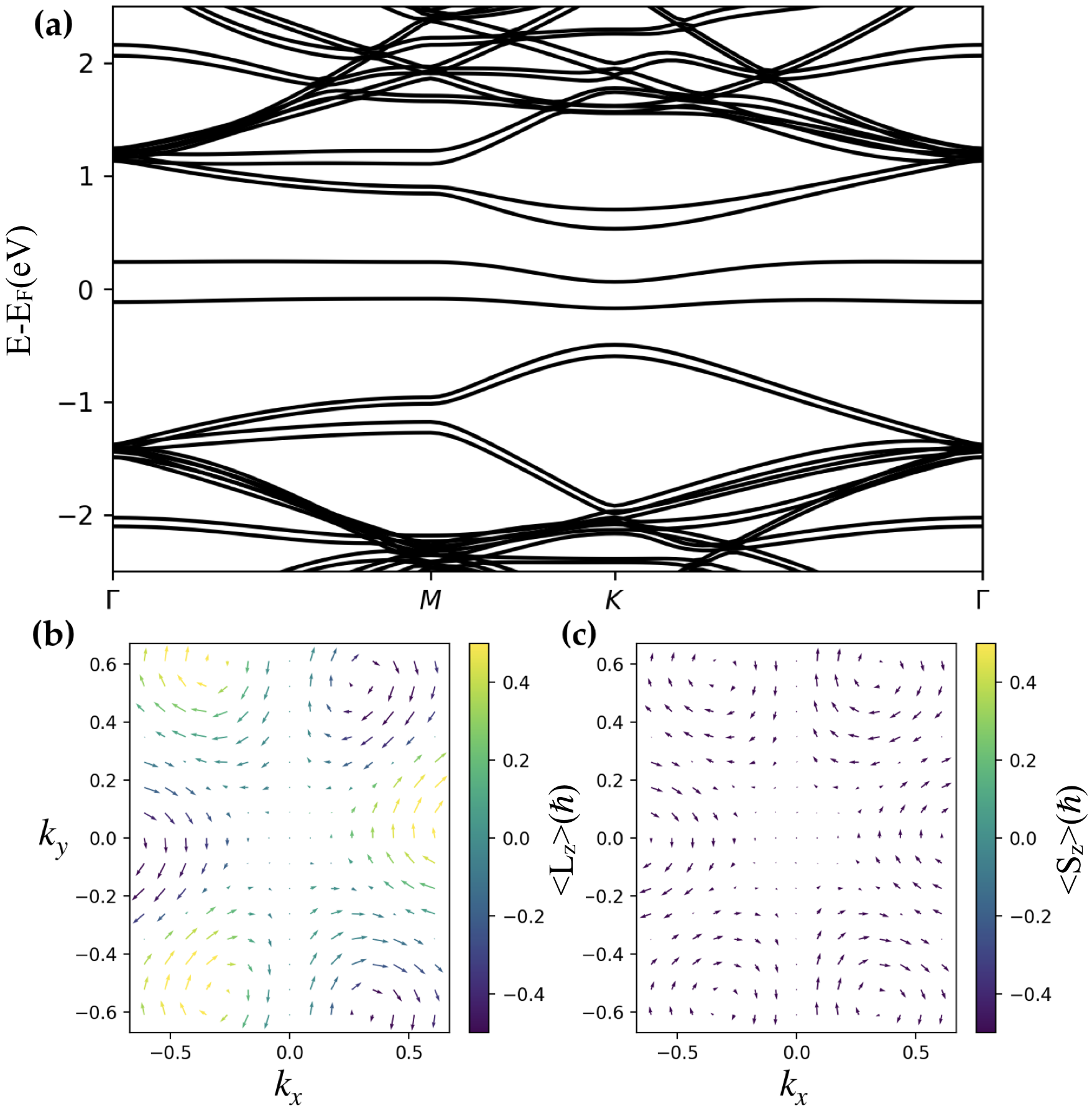}% %Here is how to import EPS art
\caption{Graphene+H electronic band structure considering spin-orbit coupling (a). Orbital texture (b) and spin texture (b) for the most energetic valence band. In this case the isolated flat bands come from the Hydrogen atom.}
\label{fig:hydrogen_graphene}
\end{figure}

We now consider graphene decorated with hydrogen. The inclusion of hydrogen is sufficient to enhance graphene's spin-orbit splitting up to 100 $\mu eV$ locally \cite{graphene_h_PhysRevLett.110.246602,Soriano2015}. For this system, we have considered a $5\times 5$ supercell with a single hydrogen atom on top at the center of the graphene flake. The band structure is presented in Fig. \ref{fig:hydrogen_graphene}(a) where spin-orbit coupling was also taken into account, showing a good agreement with previous reports \cite{Soriano2015}. The orbital and spin textures are shown in Fig. \ref{fig:hydrogen_graphene}(b) and (c) respectively, for the most energetic valence bands closer to the hydrogen states, well localized in the spectrum. Our results suggest a large imprinted $p_x$-$p_y$ hybridization which leads to the large value of the atomic orbital momentum $L_z$ having hot spots at inequivalent points in the hexagonal Brillouin zone. This is encouraging from the orbital transport perspective. Our calculations show that the intra-atomic orbital Hall conductivity (ACA) is around $0.6$ (e/2$\pi$) whereas the total (intra- and inter-atomic) orbital Hall conductivity is about $2.1$ (e/2$\pi$), which is comparable to the h-BN/graphene case discussed above. In contrast, the spin Hall conductivity is about $\approx$ $0.15$ (e/2$\pi$), still much smaller than the orbital Hall effect, but one order of magnitude larger than the spin Hall effect computed in h-BN/graphene, demonstrating the large spin-orbit coupling enhancement in this heterostructure. This result is remarkable especially considering that it solely arises from the interaction between graphene and hydrogen. On the other hand, the spin texture induces a local magnetic moment of $\sim$1 $\mu_B$, leading to a large spin splitting. The orbital Hall effect will also appear in other graphene-based heterostructures whose band structure is tuned by proximity effects \cite{Han2014,Roche2015,Pezo_2022_soc}.

\subsubsection{Germanene and bismuthene}

\begin{figure}[ht!]
\includegraphics[width=\linewidth]{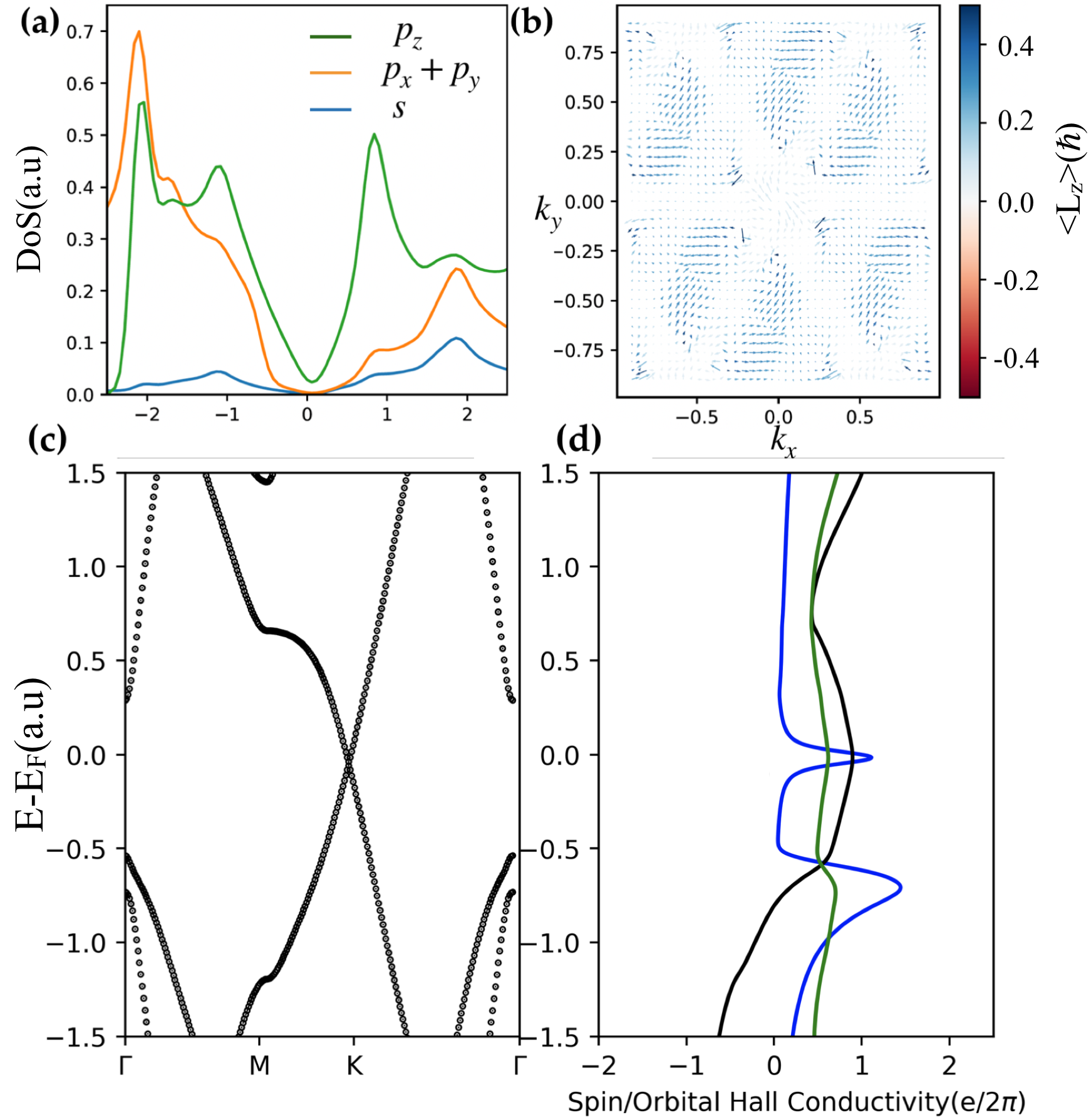}% %Here is how to import EPS art
\caption{(a) Germanene projected density of states showing the $p_z$ (green), $p_x\;,p_y$ (orange) and $s$ (blue) states as a function of the Fermi level.  (b) Corresponding orbital texture for the most energetic valence band. (c) Germanene electronic band structure, and (d) spin (blue), intra-atomic (black) and total orbital Hall conductivities (green).}
\label{fig:germanene}
\end{figure}

The next material we consider is germanene which possesses a narrow gap and has a buckled structure that favors a $sp^3$ hybridization inducing a $p_x$-$p_y$ hybridization away from the neutrality point [Fig. \ref{fig:germanene}(a)] which results in an orbital texture in momentum space [Fig. \ref{fig:germanene}(b)]. Alike Kane-Mele model with small spin-orbit coupling, germanene possesses slightly gapped Dirac cones located at K and K' points [Fig. \ref{fig:germanene}(c)]. The Hall conductivities are depicted in Fig. \ref{fig:germanene}(d) where the spin Hall effect (blue) reaches a (narrow) quantized plateau at the Fermi level, associated with the non-trivial phase. Whereas the spin Hall conductivity is peaked close to the gap, where the spin Berry curvature is maximum, the non-vanishing orbital texture in germanene leads to a finite value of the orbital Hall conductivity (intra-atomic contribution in black, total contribution in green) on a much broader range of energy around the gap. Notice that the total orbital Hall effect remains smaller than the intra-atomic Hall effect, which implies that inter-atomic and intra-atomic contributions partially cancel each other. A particular feature of germanene (and bismuthene, see below) is that inversion symmetry is preserved, and therefore the total orbital Hall response has its origin in the non-abelian nature of the Berry curvature as already shown \cite{Cysne2022}.

\begin{figure}[ht!]
\includegraphics[width=\linewidth]{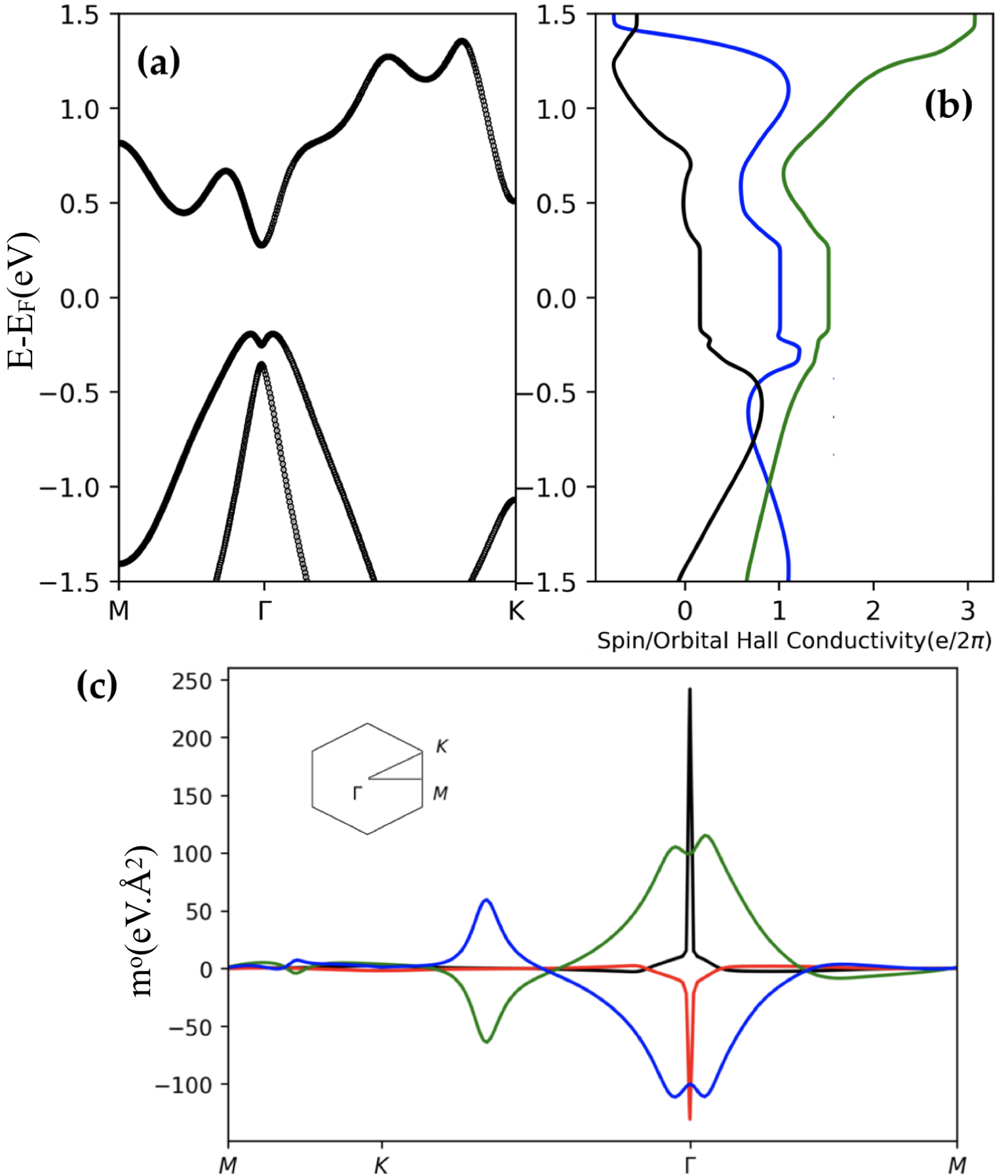}% %Here is how to import EPS art
\caption{(a) Bismuthene electronic band structure. (b) Spin (blue), intra-atomic (black) and total orbital Hall conductivities (green). (c) Orbital moment calculated for the two most (red and black) and two less energetic (blue and green) bands along the $M-K-\Gamma-M$ kpath.}
\label{fig:bismuthene_om}
\end{figure}

The last system we consider is buckled bismuthene, which displays a much larger gap than germanene due to a much larger spin-orbit coupling [Fig. \ref{fig:bismuthene_om}(a)]. In this crystalline phase, bismuthene's band character is inverted at $\Gamma$ point due to spin-orbit coupling, following the same process as described in the Bernevig-Hughes-Zhang model \cite{Bernevig1757,bismuthene_bhz}, exemplified above by the Kane-Mele model with strong spin-orbit coupling (see Fig. \ref{fig:kane_mele_soc}). This change in the band structure allows for a strong $s$-character at this point in reciprocal space leading to a quenched orbital texture \cite{Li_bism_2018}. The spin, intra-atomic and total orbital Hall conductivities are displayed on Fig. \ref{fig:bismuthene_om}(b). The absence of an orbital texture in terms of the $p_x$ and $p_y$ near the gap leads to a vanishingly smaller intra-atomic orbital Hall effect (black), which increases away from the gap due to enhanced $p_x$-$p_y$ hybridization. In contrast, the total orbital Hall conductivity reaches a large value (green), even larger than that of the spin Hall conductivity (blue). These larger values can be understood by looking at the orbital moment distribution along the momentum path $M-K-\Gamma-M$ shown in Fig. \ref{fig:bismuthene_om}(c). The hot-spots located at the $\Gamma$-point lead to a larger value of the total orbital Hall conductivity compared to the spin one. Notice that the spin conductivity is quantized, whereas the total orbital conductivity is not. We mention in passing that it has been recently suggested that such non-quantized plateaus are related to high-order topological insulating behavior \cite{Costa2022}. Hence, the overall scenario in bismuthene contrasts markedly with that in germanene and follows the situation discussed in the previous section using the Kane-Mele model. Consequently, from a materials' perspective, we are able to draw differences on the orbital response based on details of their band structures and orbital character.

\section{Impact of disorder on orbital Hall transport\label{s:3}}

An important question that remains unanswered at this point is the impact of disorder-induced scattering on the orbital conductivity. As a matter of fact, a simple-minded rationale suggests that the intra-atomic orbital Hall effect, which arises from the atomic orbital moment, would be less sensitive to momentum scattering than the inter-atomic orbital Hall effect, which arises from self-rotation of the electron wave packet in the unit cell. To investigate the impact of disorder, we consider three different systems: germanene, h-BN/graphene, the two narrow-gap semiconductors studied above, and MoS$_2$, a large band gap semiconductor that has been predicted to be an orbital Hall insulator \cite{Cysne2021,Pezo2022}. From the tight-binding basis obtained by {\it ab initio} simulations, we introduce disorder by the inclusion of an on-site Anderson disorder which can be expressed mathematically like
\begin{equation}
H=H_0+\sum_i V_i,
\end{equation}
where $H_0$ is the bare Hamiltonian corresponding to a $10\times 10$ supercell and $V_i$ is an onsite potential acting on the $i$ site with values $[-1,1]eV$. We calculated the Hall conductivity for 40 random realizations of every fixed set of parameters.

\begin{figure}
\includegraphics[width=\linewidth]{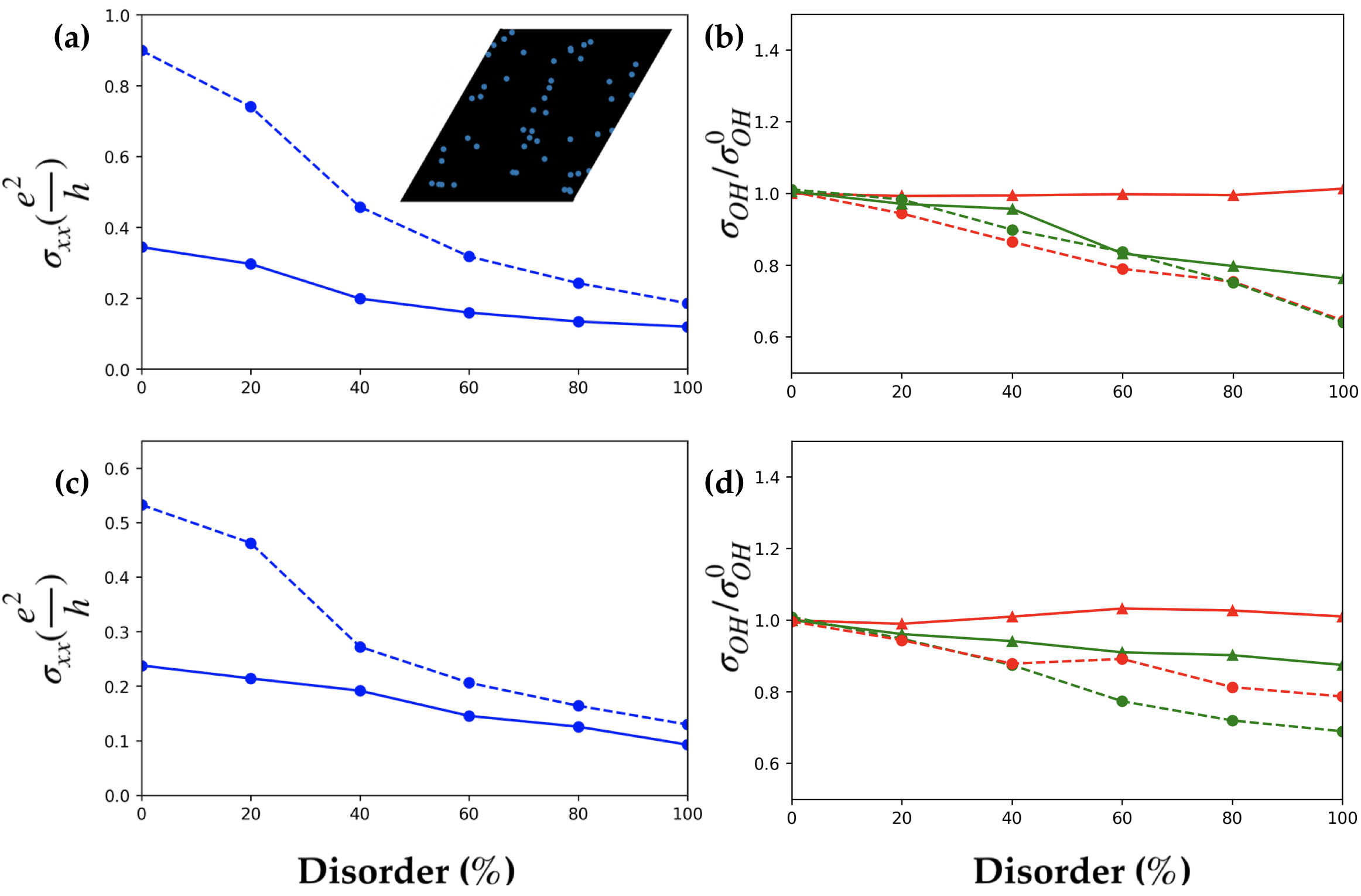}% %Here is how to import EPS art
\caption{(a,c) Longitudinal conductivity and (b,d) orbital Hall conductivity near the Fermi level ($\varepsilon=0.1$ eV) for germanene (a,b) and h-BN/graphene (c,d) as a function of the disorder concentration, the values plot correspond to the ratio orbital Hall conductivity over its maximum value $\sigma^0_{OH}$ in the pristine case. We computed both the intra-atomic (ACA - red) and total contributions (green). The inset in (a) shows a sketch of a disorder realization in the lattice.}
\label{fig:disorder_oh_germanene_hbn}
\end{figure}

The results obtained for germanene and h-BN/graphene are depicted in Fig. \ref{fig:disorder_oh_germanene_hbn}(a,b) and (c,d), respectively. To understand how disorder affects the orbital transport, we have considered two different transport regimes: (i) the single band case, where the carrier's energy to close to the gap, $\varepsilon=0.1$ eV (solid lines), and (ii) the multiband case, where the carrier's energy is far from the gap, $\varepsilon=-3.0$ eV (dashed lines). In the former, the band dispersion is mostly linear and the longitudinal conductivity of germanene (a) and h-BN/graphene (c) slowly decays as a function of disorder due to enhanced scattering. In contrast, when the energy lies far from the gap, in the multiband case (dashed), the conductivity decay is more dramatic, as expected in conventional metals. We have also computed the intra-atomic and total orbital Hall conductivities for these different situations (b,d). To better visualize the effect of disorder, we report the ratio between the Hall conductivities with and without disorder, $\sigma_{OH}/\sigma^0_{OH}$. In the single band transport regime, we find that the intra-atomic conductivity (red) is mostly flat, independent on the disorder. Nonetheless, the total orbital Hall effect (green), which contrains both intra- and inter-atomic contributions, is as a whole much more sensitive to disorder and decreases continuously. In fact, the intra-atomic Hall effect is controlled by the orbital Berry curvature of the single band and is therefore expected to be rather robust against disorder whereas the inter-atomic Hall effect, which arises from self-rotation of the wave packet in the unit cell is much more sensitive to onsite energy fluctuations brought by Anderson disorder.

In the multiband transport regime (dashed lines), we find that {\it both} the intra-atomic and total orbital Hall effect decay at a similar rate. This distinct behavior suggests that the linear dispersion of the single band transport regime has a strong impact on the robustness of the intra-atomic orbital Hall effect. In contrast, the total orbital Hall effect, that contains the inter-atomic contribution, is much more sensitive to Anderson-type disorder. We must note that the intra-atomic contribution is larger than the total one in the case of germanene while the opposite is true for h-BN/graphene. This result indicates that the inter-atomic contribution is much more sensitive to disorder than the intra-atomic one, irrespective of the transport regime.

\begin{figure}[ht!]
\includegraphics[width=\linewidth]{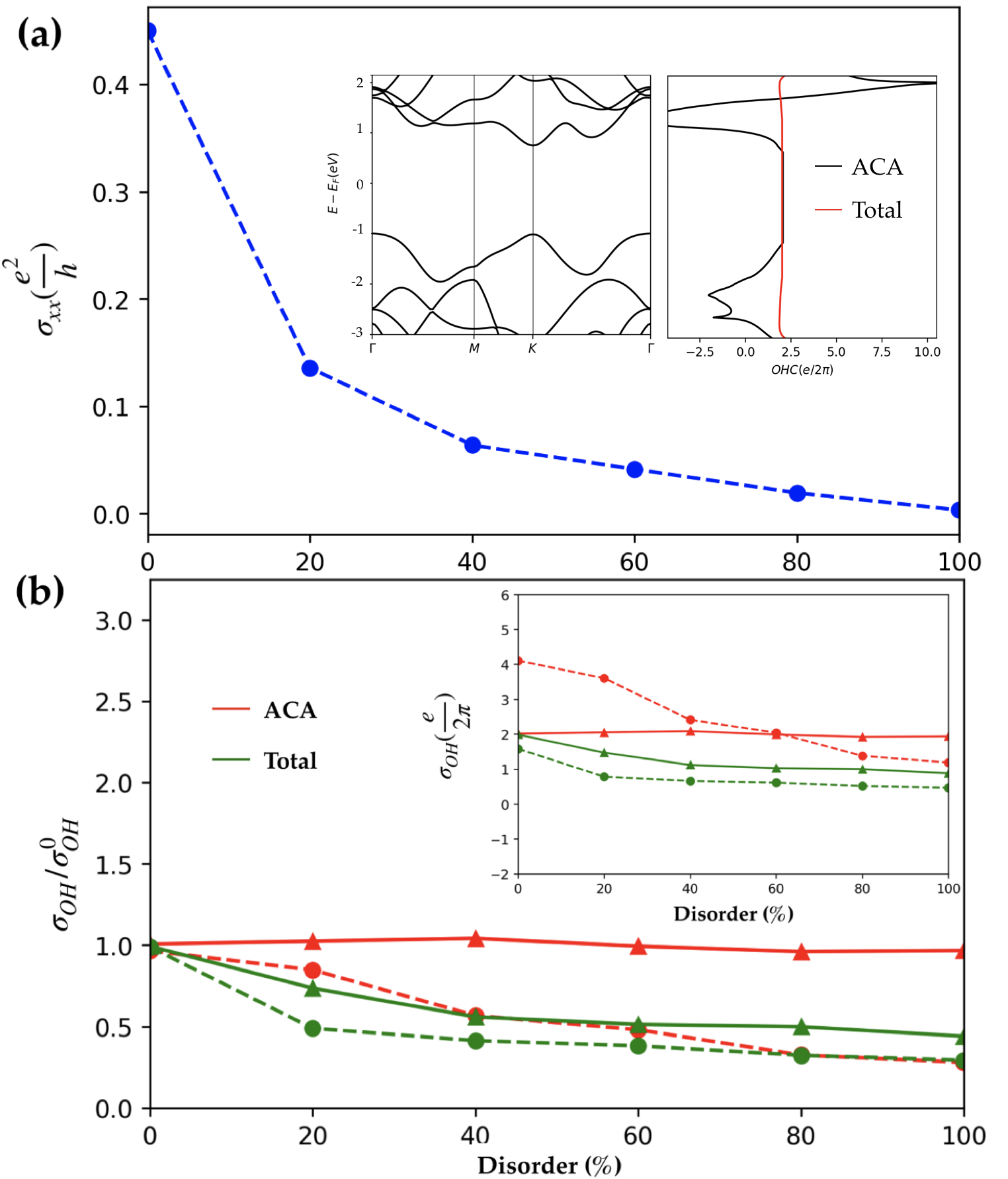}% %Here is how to import EPS art
\caption{ (a) Longitudinal conductivity for an energy of 1.5 eV with respect to the zero energy (middle of the gap) defined as the Fermi level. (b) Orbital Hall conductivities over the maximum value ($\sigma^0_{OH}$)in the pristine system, calculated by the Kubo formula in a 10$\times$10 supercell containing Anderson type disorder, intra-atomic Hall conductivity (ACA) and total orbital Hall conductivity are depicted with dashed lines for an energy of 1.5 eV with respect to the zero energy defined as the Fermi level. The solid lines correspond to the Hall conductivites calculated the Fermi level. The inset shows the actual values of the orbital conductivity for both intra-atomic and total responses.}
\label{fig:disorder_oh_xx}
\end{figure}

We now turn our attention to MoS$_2$, a large band gap orbital Hall insulator studied previously \cite{Canonico2020,Pezo2022}. The band structure and orbital Hall conductivity are reported in insert for reference. Here, we fix the carrier's energy at $\varepsilon$=1.5 eV above the center of the gap in a region where the energy dispersion is quadratic. The value of the conductivity decreases rapidly with increasing the impurity concentration, as expected in a conventional, leading to a decay over about two orders of magnitude, quite different from linearly dispersing narrow-gap germanene and graphene/hBN. The behaviour of the (normalized) orbital Hall conductivities is depicted in Fig. \ref{fig:disorder_oh_xx}(b) where dashed lines correspond to a transport energy of $\varepsilon$=1.5 eV and the solid lines correspond to a transport energy taken in the middle of the gap. The inset shows the absolute values for reference. The behavior we obtain is qualitatively similar to the one observed in the narrow-gap semiconductors discussed previously. In the gap, the intra-atomic orbital Hall conductivity (solid red) is insensitive to disorder, as expected from a Berry-curvature driven effect, whereas the total orbital Hall conductivity decays. It has been argued recently that the in-gap intra-atomic orbital Hall conductivity is associated with intra-atomic orbital polarized edge states \cite{Cysne2021,Canonico2020} that remain insensitive to the disorder, although this picture might change when considering open boundary conditions like in nanoribbons for instance \cite{C9CP01590F}. This contrasts to what we find for the total Hall response where a drop of nearly half of the initial value is observed. When the energy is set in the conduction band, one finds that both intra-atomic and total orbital Hall conductivities decrease with a similar rate, as already observed in narrow-gap semiconductors. 

To improve our study of the disorder we have considered $30 \times 30$ supercells  of a two-dimensional model system representing $d_{xy}-d_{x^2-y^2}$ and $p_z$ orbitals for a 3-band MoS$_2$ \cite{PhysRevB.88.085433}, capable to display both intra-atomic and inter-atomic orbital Hall effects (not shown). We were able to retrieve similar results calculated using the full Wannier Hamiltonian: in the gap, the total orbital Hall response decreases upon increasing disorder while the intra-atomic orbital Hall effect is preserved. These calculations lead to qualitatively the same conclusions as for the system with a larger number of orbitals. Our simulations therefore suggest that the total orbital Hall contribution is more sensitive to disorder and corroborates our previous comment regarding its relation with the periodic functions living in the bulk.

The present study, based on a real-space random potential distribution, in principle covers extrinsic scattering mechanisms such as side-jump and skew-scattering, which are known to be central to spin and anomalous Hall effects \cite{Nagaosa2010}. Indeed, the inclusion of disorder leads to a drop of the orbital Hall conductivity for states in the valence band, which seems to corroborate the conclusion drawn in \cite{PhysRevB.102.155302}. In this work, it is shown that for a 2-band model system the side-jump contribution totally cancels the Berry-curvature contribution. Furthermore, our results exhibit a similar behavior for both the intra-atomic and the total orbital Hall contributions in the metallic regime.

%Our study of the impact of disorder on orbital transport, which was performed using a real space random potential distribution, in principles covers both types of side-jump and skew-scattering mechanisms. Indeed, the inclusion of disorder leads to a drop of the orbital Hall conductivity for states in the valence band, which seems to corroborate the conclusion drawn in \cite{PhysRevB.102.155302}. In this work, it is shown that for a 2-band model system the side-jump contribution totally cancels the Berry-curvature contribution. Furthermore, our results exhibit a similar behavior for both the intra-atomic and the total orbital Hall contributions in the metallic regime.}

\section{Conclusion\label{s:4}}

In summary, we have explored the microscopic origin of the orbital Hall effect in model and realistic two-dimensional Dirac materials. Since the orbital Hall effect is intimately connected with the Berry curvature of the material, we first investigated the inter-atomic orbital Hall contribution in the Kane-Mele model, that accommodates topological phase transition and in which the intra-atomic contribution is absent. We found that although the orbital moment itself behaves differently in the topologically trivial and non-trivial phases, the resulting orbital Hall conductivity is rather controlled by the size of the gap, irrespective of its topological nature.\par

We then studied the orbital Hall effect in selected two-dimensional materials, starting with graphene. Whereas orbital Hall effect is absent of pristine graphene, it can be turned on by inducing a global or local gap, either interfacing graphene with h-BN or by using hydrogen adatoms, respectively. In these cases, the emergence of orbital and spin textures in reciprocal space stands out as a key ingredient for the generation of orbital and spin polarized Hall currents. These predictions are particularly intriguing given that these two systems are made out of light elements unable to portray a sizeable spin-orbit coupling by their own. Although the experimental distinction between entangled spin and orbital signals remains a challenge even in materials with moderate spin-orbit coupling, our results are encouraging as they show that light materials with negligible spin orbit coupling would display a large orbital Hall effect.

We then moved on to investigate the orbital Hall currents in two selected two-dimensional topological insulators, germanene and bismuthene, which represent two distinct realizations of the Kane-Mele model, with weak and strong spin-orbit coupling, respectively. In germanene, we have found that the intra-atomic orbital Hall contribution displays a larger value than of the total orbital Hall one, resembling the weakly spin-orbit coupled non-trivial phase of the Kane-Mele model (small spin-orbit gaps at K and K' points). The existence of the quantum spin Hall effect in germanene is corroborated with a narrow plateau appearing for the spin Hall conductivity while the values for the orbital Hall effect remains larger in a broader energy window. In bismuthene, besides the spin Hall effect, we have found a large orbital Hall conductivity coming from the orbital moment carried out by the bands near the Fermi level. This situation resembles the strongly spin-orbit coupled non-trivial phase of the Kane-Mele model (large spin-orbit gap at M point). 

Finally, we investigated the impact of disorder on the intra-atomic and inter-atomic contributions of the orbital Hall effect in two-dimensional systems featuring very different transport regimes (insulating, single-band and multiband metallic regimes). We find that the intra-atomic orbital Hall effect tends to be less affected by disorder than the total orbital Hall effect, especially in the insulating and single-band regimes, i.e., in situations where the orbital Berry curvature is smooth and well-defined. In contrast, in the multiband transport regime, both intra-atomic contribution and total orbital conductivity. These results suggest that irrespective of the transport regime, the intra-atomic part of the orbital Hall effect is more robust than the inter-atomic part.

The present work sheds light on the mechanisms responsible for orbital Hall effect in two-dimensional materials, and in particular clarifies the role of the gap. The intimate connection between the orbital Hall transport and the Berry curvature of the band structure opens interesting perspectives for the external control of the orbital transport through interfacial engineering or strain, as demonstrated in Ref.\cite{Son2019}. From this standpoint Van der Waals heterostructures made of light materials, such as graphene and h-BN for instance, could be used for the realization of nonlocal orbital devices, akin to the all-electric valley or spin Hall transistor \cite{Gorbachev2014,Choi2018}. In this context, a more comprehensive understanding of the orbital relaxation induced by momentum scattering is necessary.

\section{Acknowledgments}
This work was supported by the ANR ORION project, grant ANR-20-CE30-0022-01 of the French Agence Nationale de la Recherche. D. G.O. and A. M. acknowledge support from the Excellence Initiative of Aix-Marseille Universit\'e - A*Midex, a French "Investissements d'Avenir" program.

\bibliography{refs-resub-2}

\end{document}